\documentclass[aps,nofootinbib,preprint]{revtex4}
\usepackage{amsmath,amssymb}
\usepackage{graphicx}

\def\bi{\boldsymbol i}
\def\bq{\boldsymbol{q}}

\begin{document}

%%%%%%%%%%%%%%%%%%%%%%%%%%%%%%%%%%%%%%%%%%%%%%%%%%%%%%%%%%%%%%
\title{
Embedding the Cremmer-Scherk Configuration into SO(16) and Effective SO(10) Gauge Symmetry
}

\date{\today}

\author{Hironobu Kihara\footnote{hkihara(at)th.phys.titech.ac.jp}}
\affiliation{Faculty of Science and Interactive Research Center of Science, 
Graduate School of Science and Engineering,
 Tokyo Institute of Technology, 2-12-1 Oh-okayama, Meguro, Tokyo 152-8551, Japan\\
The Faculty of Business and Commerce, Keio University, 2-15-45, 
Mita, Minato ku, Tokyo 108-8345, Japan\\
Research and Education Center for Natural Sciences,
Keio University,\\ 
4-1-1 Hiyoshi, Yokohama,
Kanagawa 223-8521, Japan\\
Faculty of Science and Technology, Seikei University, 
3-3-1, Kichijoji-Kitamachi, Musashino-shi, Tokyo 180-8633, Japan
}

%\preprint{TIT/HEP-603}

\begin{abstract}
We show the explicit embedding of the Cremmer-Scherk configuration into SO(16) Gauge theory and 
that the non-Abelian flux breaks the gauge symmetry SO(16) to SO(10).  
Adjoint scalar fields of SO(10) coming from components of six compact directions become massive. 
There are several scalar fields beloging to the representation {\bf 10}, which may break SO(10). 
\end{abstract}

\maketitle
%%%%%%%%%%%%%%%%%%%%%%%%%%%%%%%%%%%%%%%%%%%%%%%%%%%%%%%%%%%%%%%%%%%%

Recently we studied the numerical solution of the dynamical compactification  the ten-dimensional SO(6)
Einstein-Yang-Mills theory \cite{Kihara:2009ea}, by using the Cremmer-Scherk configuration on $S^6$ \cite{Cremmer:1976ir}. 
The gauge symmetry SO(6) is completely broken by the non-Abelian flux \cite{Chingangbam:2009jy}. 
We discussed the possibility to embed the Cremmer-Scherk configuration into larger groups \cite{Kihara:2009gc}. 
In  \cite{Chingangbam:2009jy}, the symmetry breaking was realized without any additional ten-dimensional Higgs fields. 
The extra-dimensional components of gauge fields behave like Higgs fields in four-dimensional effective theory. 

In this note we embed the Cremmer-Scherk configuration on $S^6$ into SO(16) explicitly. 
To consider the group SO(16) is inspired by \cite{Horava:1995qa}. 
We expect that the gauge symmetry SO(16) breaks to SO(10) on the Cremmer-Scherk configuration. 
The topological discussion in \cite{Kihara:2009gc} is for the existence of the suitable tensor $K$ shown later. 

We work on the ten-dimensional SO(16) Einstein-Yang-Mills theory with Tchrakian's higher derivative coupling term of Yang-Mills field strength \cite{Tchrakian:1978sf}. This is same as the model in \cite{Kihara:2009ea} except for the gauge group. 
The space-time is the dynamically compactified product of the four-dimensional Friedmann-Lemaitre-Robertson-Walker universe and 
a six-dimensional sphere, 
$FLRW_4 \times_{\rm DC} S^6$,  shown in \cite{Kihara:2009ea}. 
Coordinates on the $FLRW_4$ are represented by $x^{\mu}~ (\mu=0,1,2,3)$ and 
coordinates on the sphere $S^6$ are by $y^i ~(i=4,5,\cdots,9)$. 
Suppose that the six-dimensional sphere shrinks to a constant radius and the constants, $G, \alpha, q, V_0$,
 satisfy the condition for the nonexistence of the tachyonic fluctuations.

Let us start from Clifford algebra. 
We use three Clifford algebras with respect to SO(6), SO(10) and SO(16). 
$\gamma_a$ are $8 \times 8$ matrices and $\tilde{\gamma}_{\alpha}$ are 
$32 \times 32$ matrices. These matrices satisfy the following anti-commutation relations, 
\begin{align}
\{ \gamma_a , \gamma_b \} &= 2 \delta_{ab}~~~~(a,b=1,2,\cdots ,6)~,\cr
\{ \tilde{\gamma}_{\alpha} , \tilde{\gamma}_{\beta} \} &= 
2 \delta_{\alpha \beta} ~~~~(\alpha , \beta =1,2,\cdots ,10) ~.
\end{align}
Chirality matrices are defined as follows, 
\begin{align}
\gamma_7 &:= - i \gamma_1 \cdots \gamma_6 ~,\cr
\tilde{\gamma}_{11} &:= i \tilde{\gamma}_1 \cdots \tilde{\gamma}_{10} ~.
\end{align}
$\Gamma_{\underline{A}} ~(\underline{A}=1,2,\cdots,16)$ are constructed by tensor products of $\gamma_a$ and $\tilde{\gamma}_{\alpha}$, 
\begin{align}
\Gamma_1 &= \gamma_1 \otimes {\bf 1}_{32} ~,\cr
\vdots & \cr
\Gamma_6 &= \gamma_6 \otimes {\bf 1}_{32} ~,\cr
\Gamma_7 &= \gamma_7 \otimes \tilde{\gamma}_1 ~,\cr
\Gamma_8 &= \gamma_7 \otimes \tilde{\gamma}_2 ~,\cr
\vdots & \cr
\Gamma_{16} &= \gamma_7 \otimes \tilde{\gamma}_{10} ~.
\end{align}
Here the index $\underline{A}$ is used for labeling not the ten-dimensional space-time coordinates 
but the internal directions.
The chirality matrix with respect to the group SO(16) is defined as follows, 
\begin{align}
\Gamma_{17} &:= \Gamma_{1} \cdots \Gamma_{16} \cr
&= \gamma_1 \cdots \gamma_6 \otimes \tilde{\gamma}_1 \cdots \tilde{\gamma}_{10} \cr
&= \gamma_7 \otimes \tilde{\gamma}_{11}
\end{align}
We use the following matrix, $K$, 
\begin{align}
K &:= \gamma_7 \otimes {\bf 1}_{32}~.
\end{align}
The infinitesimal generators, $\gamma_{ab}$, $\tilde{\gamma}_{\alpha\beta}$,
 and $\Gamma_{AB}$ are defined as follows, 
\begin{align}
 \gamma_{ab} &:= \frac{1}{2} [ \gamma_a , \gamma_b  ]  ~,~~~~(a,b = 1,2, \cdots , 6)\cr
\tilde{\gamma}_{\alpha\beta}&:= \frac{1}{2} [ \tilde{\gamma}_{\alpha} , \tilde{\gamma}_{\beta}  ] 
 ~,~~~~(\alpha,\beta = 1,2, \cdots , 10)\cr
\Gamma_{\underline{A},\underline{B}} &:= \frac{1}{2} [\Gamma_{\underline{A}} , \Gamma_{\underline{B}}]
 ~,~~~~(\underline{A},\underline{B} = 1,2, \cdots , 16) ~.
\end{align}
Generators $\Gamma_{\underline{A},\underline{B}}$ are explained in terms of $\gamma_{ab}$, $\tilde{\gamma}_{\alpha\beta}$ and so on,
\begin{align}
\Gamma_{ab} &= \gamma_{ab} \otimes {\bf 1}_{32} ~, \cr
\Gamma_{a,\alpha+6} &= \frac{1}{2} \left( \gamma_a \gamma_7 
 -  \gamma_7  \gamma_a  \right) \otimes \tilde{\gamma}_{\alpha} ~,\cr
\Gamma_{7+\alpha , 7 + \beta} &= {\bf 1}_8 \otimes \tilde{\gamma}_{\alpha\beta} ~.
\end{align}
Hence $\Gamma_{ab}$ and $\Gamma_{7+\alpha,7+\beta}$ commute with each other. 
$\Gamma_{ab}$ commute with $K$, too. 
\begin{align}
[  \gamma_{ab} \otimes {\bf 1}_{32} , K ] &=0~.
\end{align}
The Cremmer-Scherk configuration in the group SO(16) is obtained as, 
\begin{align}
A^{(0)} = \frac{1}{ 4 \bq R_2} \Gamma_{ab} y^{a} V^{b} ~, \quad
F^{(0)} = \frac{1}{ 4 \bq R_2^2} \Gamma_{ab} V^{a} \wedge V^{b}  ~.
 \label{eq:sol-R} 
\end{align}
This configuration satisfies the following self-duality equation, 
\begin{align}
F^{(0)} &=  *_6 \bi  \frac{\bq R_2^2}{3}  
K F^{(0)} \wedge F^{(0)}  ~,
\label{eq:sde}
\end{align} 
Here $K$ is covariantly constant on the background $A^{(0)}$, 
\begin{align}
D^{(0)} K &= d K + \bq [ A^{(0)} ,  K  ] = 0 ~, 
\end{align}
because $K$ commutes with $A^{(0)}$. Such a covariantly constant tensor is required in \cite{Kihara:2009gc}. 
The fluctuation $\delta A$ split up into two parts as shown in \cite{Chingangbam:2009jy}. 
\begin{align}
A &= A^{(0)} + \delta A~,\cr 
\delta A &=  v + \Phi ~,\cr
v & = \frac{1}{2} v_{\mu}^{\underline{A},\underline{B}} \Gamma_{\underline{A},\underline{B}} dx^{\mu}~~~~(\mu = 0,1,2,3)~,\cr
\Phi &= \frac{1}{2} \Phi_i^{\underline{A},\underline{B}} \Gamma_{\underline{A},\underline{B}} dy^i~~~~~(i=4,5,\cdots, 10) ~.
\end{align}
These components are decomposed with respect to representations under the gauge transformation SO(10), 
\begin{align}
\left( v_{\mu}^{\underline{A},\underline{B}} \right) &= 
\begin{pmatrix}  a_{\mu}^{\alpha \beta} & z_{\mu}^{\alpha a} \\
 - z_{\mu}^{\alpha a} & w_{\mu}^{ab}  \end{pmatrix} ~,\cr
\left( \Phi_i^{\underline{A},\underline{B}} \right) &= 
\begin{pmatrix}  h_i^{\alpha \beta} & k_{i}^{\alpha a} \\
 - k_{i}^{\alpha a} & \varphi_{i}^{ab}  \end{pmatrix} ~.
\end{align}
%In the case of SO(6) Yang-Mills theory, $v$ becomes massive four-dimensional vector field. 
The mass terms of effective four-dimensional vector fields are yielded from terms including 
$D^{(0)}v = d^{(6)} v + q [ A^{(0)} , v ]$. Therefore 
it turns out that the lowest Kaluza-Klein modes of any fluctuations $v$, which satisfy $[ A^{(0)} , v]=0$ 
become massless. Hence effective four-dimensional vector fields, $a_{\mu}^{\alpha \beta}(x)$, which takes values in SO(10) 
are massless. While SO(6) in SO(16) is completely broken same as in \cite{Chingangbam:2009jy}. 
In other words, $ w_{\mu}^{ab}(x)$ are massive. 
Hence we obtain the symmetry breaking SO(16) $\rightarrow$ SO(10). 
One may imagine that the lowest Kaluza-Klein modes of the scalar fields, $h_i^{\alpha \beta}(x,y)$, 
which take values in SO(10) become massless. 
Because six-dimensional sphere does not have nontrivial harmonic one-forms, 
the lowest Kaluza-Klein modes of $h_i^{\alpha \beta}(x,y)$ are not massless.

 Four-dimensional vector fields, $z_{\mu}^{\alpha a}(x,y)$ ($a=1,2,\cdots ,6$), 
and scalar fields, $k_{i}^{\alpha a}(x,y)$ $(a,i =1,2,\cdots, 6)$,
 belong to the representation {\bf 10} 
under the gauge group SO(10). 
Naively, the lowest Kaluza-Klein modes of these fields have nonvanishing cross terms with $F^{(0)}$ and 
 get mass by the Cremmer-Scherk configuration. 
$h_i^{\alpha \beta}(x,y)$ and $k_{i}^{\alpha a}(x,y)$ are candidates of 
Higgs fields breaking SO(10) to SU(3)$\times$SU(2)$\times$U(1).

\section*{Acknowledgments}
The work of H.K.~is supported by the Interactive Research Center of Science, 
Graduate School of Science and Engineering,
 Tokyo Institute of Technology.

%%%%%%%%%%%%%%%%%%%%%%%%%%%%%%%%%%%%%%%%%%%%%%%%%%%%%%%%%%%%%%%%%%%%%%%%%%%%%

\end{document}